\documentclass[12pt]{article}

\usepackage{osajnl}
\usepackage{overcite,hyperref}
\usepackage{times}
\usepackage{graphicx} 
\usepackage{amsmath} 
\usepackage{bbm}

\newtheorem{remark}{Remark}

\begin{document}

\title{Point target detection and subpixel position estimation in
optical imagery}


\author{Vincent Samson, Fr\'ed\'eric Champagnat}
\affiliation{Office National d'\'Etudes et de Recherches A\'erospatiales,
\\ 29, avenue de la division Leclerc, 92322 Ch\^atillon Cedex, France}
\author{Jean--Fran\c{c}ois Giovannelli}
\affiliation{Laboratoire des Signaux et Syst\`emes,
\\ Sup\'elec, Plateau de Moulon, 91192 Gif--sur--Yvette Cedex, France}

\begin{abstract}
This paper addresses the issue of detecting point objects in a
clutter background and estimating their position by image processing.
We are interested in the specific context where the object signature
significantly varies with its random subpixel location because of
aliasing. Conventional matched filter neglects this phenomenon and
causes consistent loss of detection performance.  
Thus, alternative detectors are proposed and numerical results show 
the improvement brought by approximate and generalized likelihood ratio tests 
in comparison with pixel matched filtering.
We also study the performance of two types of subpixel position estimators. 
Finally, we put forward the major influence of sensor design on both estimation 
and point object detection performance.
\end{abstract}


\maketitle


\section{Introduction}
\label{sec:intro}

\begin{figure}[ht]
\begin{center}
\includegraphics[height=4.5cm]{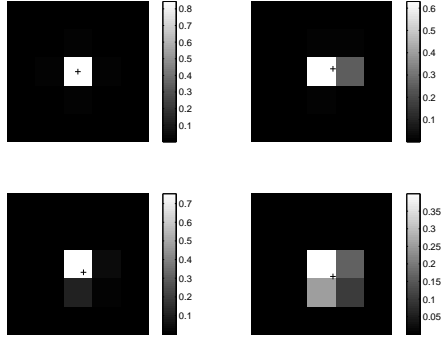}
\caption[]{Examples of image spots for different cross-marked subpixel positions (windows of size $5 \times 5$ pixels). Sensor design parameter $r_c$ is set to its common value of $2.44$ (see section~\ref{sec:application}).} 
\label{fig:taches}
\end{center}
\end{figure}

\begin{figure}[ht]
\begin{center}
\includegraphics[height=4cm]{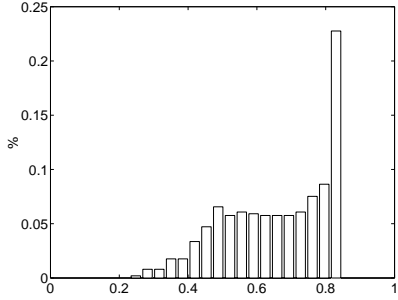}
\caption[]{Empirical distribution of the image-spot central pixel $\boldsymbol{s}_{\boldsymbol{\epsilon}}[0,0]$ for a uniformly random position $\boldsymbol{\epsilon} \sim \mathcal{U}_{[-0.5,0.5[^2}$.}  
\label{fig:histo}
\end{center}
\end{figure}

We tackle the problem of subpixel object detection in image sequences
which arises for instance in infrared search and track (IRST)
applications. In this context, the target signature is proportional to:
\begin{equation}
\boldsymbol{s}_{\boldsymbol{\epsilon}}[i,j] = \int_{i-0.5}^{i+0.5}
\int_{j-0.5}^{j+0.5} h_o(u-\epsilon_1,v-\epsilon_2) \, du \, dv.
\label{eq:signal}
\end{equation}
$\boldsymbol{s}_{\boldsymbol{\epsilon}}[i,j]$ represents the
percentage of light intensity at pixel $(i,j)$,
$\boldsymbol{\epsilon}=(\epsilon_1,\epsilon_2)$ refers to the object
random subpixel position and $h_o$ is the optical point spread
function (PSF).  According to common sensor design, the energy of the
signal component $\boldsymbol{s}=\alpha
\boldsymbol{s}_{\boldsymbol{\epsilon}}$ is almost concentrated on a
single pixel.  However, contrary to the amplitude $\alpha$ which is
unknown too, dependence on the location parameter
$\boldsymbol{\epsilon}$ is highly nonlinear.   Its influence in our
application is rather significant because of aliasing and unless a
velocity model is available, object subpixel position is hardly
predictable from frame to frame.   Actually, common sensor design
leads to an image spot downsampled by almost a factor 5.   We can see
on Figure~\ref{fig:taches} the energy loss at central pixel according
to subpixel location and the random change in spatial pattern due to
aliasing.
This phenomenon has a major impact on detection performance as shown
thereafter.  To our knowledge, this pitfall has not been addressed yet
in the literature.   A prevailing opinion stands that there is no
signature information in subpixel objects.  Indeed, the different
authors dealing with small object detection have concentrated on
clutter removing \cite{Wang82,Margalit85,Soni93}, multi- or
hyper-spectral fusion \cite{Yu97,Ashton98} and multiframe tracking
methods \cite{Reed83,Blostein91,Mooney95}.    We focus here on the
processing of a single frame.   In section~\ref{sec:problem}, we
formulate the detection problem in the classical model of a signal in
additive Gaussian noise \cite[ch.2-4]{VanTrees68}.  When the signal is
deterministic, Neyman-Pearson strategy yields the conventional matched
filter.  In the present case, the signal from the target depends on
unknown parameters and we have to deal with a composite hypothesis
test.  A common procedure is given by the generalized likelihood ratio
test.   But the ``\,nuisance\,'' parameters $\alpha$ and
$\boldsymbol{\epsilon}$ can also be considered as random variables
with known distributions (some \textit{a priori} density functions in
the Bayesian terminology), then the straightforward extension of the
likelihood ratio test is to integrate the conditional distribution
over $\alpha$ and $\boldsymbol{\epsilon}$.  When modelling the signal
component as a sample function, we could also think of the class of
random signal in noise detection problems, which have essentially been
studied in the Gaussian case.  Unfortunately, considering
$\boldsymbol{s}_{\boldsymbol{\epsilon}}$ as a random vector, its
empirical distribution proves to be highly non Gaussian when
$\boldsymbol{\epsilon}$ is uniformly sampled.  For instance, the
histogram of the central pixel depicted on Figure~\ref{fig:histo}
shows that a Gaussian fit is not satisfactory at all.
In section~\ref{sec:application}, we define more precisely the optical
system model used in our numerical experiments. We consider both a
Gaussian white noise and a fractal noise of unknown correlations
generated by a standard technique of spectral synthesis.
Section~\ref{sec:position-estimation} is devoted to the position
estimation problem, i.e. estimation of parameter
$\boldsymbol{\epsilon}$. We propose two estimators that take
into account the fact that the signal amplitude $\alpha$ is also
unknown.   We demonstrate the performance of these estimators in terms
of mean square errors.   As for the detection problem, we finally
illustrate the expected improvement in quality brought by a correctly
sampled optics compared to common sensor design.

\section{Detection problem}
\label{sec:problem}

We consider a local detection window sliding across the image. The
problem is to decide whether an object is present or not at the window
central pixel.  This is a binary test which typically reads
as follows:
\begin{equation}
\left\{
\begin{array}{lcl}
\mathrm{H_0}&:&\boldsymbol{z}=\boldsymbol{n} \\
\mathrm{H_1}&:&\boldsymbol{z}=\alpha
\boldsymbol{s}_{\boldsymbol{\epsilon}}+\boldsymbol{n}
\end{array}
\right.
\label{eq:test}
\end{equation}
where $\boldsymbol{z}$ is the vector collecting the window data,
$\boldsymbol{s}=\alpha \boldsymbol{s}_{\boldsymbol{\epsilon}}$ is the
object response (signal vector) and $\boldsymbol{n}$ the additive
Gaussian noise.   The signature shape is known and deterministic, so
that $\boldsymbol{s}$ only depends on the two unknown parameters
$\alpha \in \mathbbm{R}$ and $\boldsymbol{\epsilon} \in
\mathcal{E}=[-0.5,0.5[^2$.   The noise vector $\boldsymbol{n}$ is
supposed to be centered (in practice we first remove the empirical
mean from the data) with a known or previously estimated covariance matrix $\boldsymbol{R}$.   
Thus, if we assume that $\boldsymbol{n}$ is independent from $\boldsymbol{s}$, the
following conditional distributions are Gaussian:
\begin{equation}
\left\{
\begin{array}{lcl}
p(\boldsymbol{z} | \mathrm{H_0}) &\sim& \mathcal{N}(0,\boldsymbol{R})
\\ p(\boldsymbol{z} | \mathrm{H_1},\alpha,\boldsymbol{\epsilon})
&\sim& \mathcal{N}(\alpha
\boldsymbol{s}_{\boldsymbol{\epsilon}},\boldsymbol{R})
\end{array}
\right.
\label{eq:conditionaldist}
\end{equation}

Let first assume that parameters $\alpha$ and $\boldsymbol{\epsilon}$
are given. The problem amounts to a simple hypothesis test which is to
detect a deterministic signal in a Gaussian noise. 
The Neyman-Pearson strategy or likelihood ratio test (LRT) is given by:
\begin{equation}
\begin{array}{lcl}
\displaystyle
\frac{p(\boldsymbol{z}|\mathrm{H_1},\alpha,\boldsymbol{\epsilon})}{p(\boldsymbol{z}|\mathrm{H_0})}
\begin{array}{c}
\mathrm{H_1}\\ > \\ < \\ \mathrm{H_0}
\end{array}
\mathrm{threshold}
\end{array}
\label{eq:lrt}
\end{equation}
It is equivalent to classical matched filtering which simply
compares the statistic  $\alpha
\mathcal{T}_{\boldsymbol{\epsilon}}(\boldsymbol{z})=\alpha \,
\boldsymbol{s}_{\boldsymbol{\epsilon}}^t \boldsymbol{R}^{-1}
\boldsymbol{z}$ with some threshold.

\subsection{Pixel matched filtering}
\label{subsec:matched-filtering}

\begin{figure}[ht]
\begin{center}
\includegraphics[height=4.5cm]{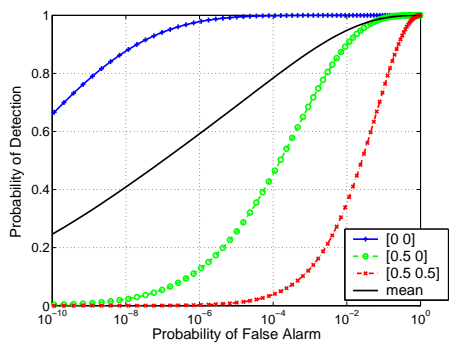}
\caption[]{Examples of pixel matched filter theoretical ROC curves for $\boldsymbol{\epsilon}^{\star}=\boldsymbol{\epsilon}_0$ in blue (ideal curve), $\boldsymbol{\epsilon}^{\star} \neq \boldsymbol{\epsilon}_0$ in green and red (worst case), and finally the mean curve in black for uniformly sampled $\boldsymbol{\epsilon}^{\star}$ (SNR $=15$dB).}  
\label{fig:corthq15db}
\end{center}
\end{figure}

The exact object location being unknown in practice, we could assume
by default that $\boldsymbol{\epsilon}=\boldsymbol{\epsilon}_0=[0,0]$,
i.e. the object is at the center of the pixel, whereas
the true location would correspond to
$\boldsymbol{\epsilon}=\boldsymbol{\epsilon}^{\star}$.   Thus, the
detector which consists in thresholding the pixel matched filter (PMF)
$\alpha \mathcal{T}_{\boldsymbol{\epsilon}_0}(\boldsymbol{z})$ is
optimum provided
$\boldsymbol{\epsilon}^{\star}=\boldsymbol{\epsilon}_0$.   Otherwise
it is mismatched and therefore suboptimum.    Since conditional
distributions of
$\mathcal{T}_{\boldsymbol{\epsilon}_0}(\boldsymbol{z})$ under each
assumption are Gaussian, we easily get the expression of the
probalility of detection $P_d$ and of false alarm $P_{fa}$. Corresponding
receiver operating characteristic (ROC) curves for critical values of
$\boldsymbol{\epsilon}^{\star}$  are depicted on
Figure~\ref{fig:corthq15db}. They clearly show that the PMF
performances are significantly worse as $\boldsymbol{\epsilon}_0$
differs from $\boldsymbol{\epsilon}^{\star}$.  But beyond extreme
situations (related to a true target location between two or four
pixels instead of the center), the ``\,mean curve\,'' represents the
average statistics over uniformly random positions.  Compared to the
ideal curve, we can see that the price paid if one neglects the random
location is rather high even at favorable signal-to-noise ratio. For a
SNR of $15$dB and at a $P_{fa}$ of $10^{-4}$,
probability of detection decreases from nearly $1$ to $0.8$.

The object response also depends (linearly this time) on the amplitude
$\alpha$, which is generally unknown.   Yet, assuming strictly
positive amplitude, we see that whatever $\alpha > 0$, thresholding
$\alpha \mathcal{T}_{\boldsymbol{\epsilon}_0}(\boldsymbol{z})$ gives
the same ROC curve as thresholding
$\mathcal{T}_{\boldsymbol{\epsilon}_0}(\boldsymbol{z})$.   Without any
assumption on $\alpha$, a classical solution is to estimate it by
maximum likelihood (ML).   Indeed under the Gaussian noise assumption,
the optimum in $\alpha$ for a given $\boldsymbol{\epsilon}$ is
explicit:
\begin{eqnarray}
\hat{\alpha}(\boldsymbol{\epsilon}) &=& \displaystyle \arg
\max_{\alpha \in \mathbbm{R}} \; p(\boldsymbol{z} |
\mathrm{H_1},\alpha, \boldsymbol{\epsilon}) \nonumber \\ &=&
\displaystyle \arg \min_{\alpha \in \mathbbm{R}} \;
\left\{(\boldsymbol{z}-\alpha
\boldsymbol{s}_{\boldsymbol{\epsilon}})^t \boldsymbol{R}^{-1}
(\boldsymbol{z}-\alpha \boldsymbol{s}_{\boldsymbol{\epsilon}})
\right\} \nonumber \\ &=& \displaystyle
\frac{\boldsymbol{s}_{\boldsymbol{\epsilon}}^t \boldsymbol{R}^{-1}
\boldsymbol{z}}{\boldsymbol{s}_{\boldsymbol{\epsilon}}^t
\boldsymbol{R}^{-1} \boldsymbol{s}_{\boldsymbol{\epsilon}}}
\label{eq:alphaml}
\end{eqnarray}
and then the ``\,generalized\,'' pixel matched filter (referred to as
GPMF) is equal to
\begin{equation}
\hat{\alpha}(\boldsymbol{\epsilon}_0)
\mathcal{T}_{\boldsymbol{\epsilon}_0}(\boldsymbol{z}) = \displaystyle
\frac{|\boldsymbol{s}_{\boldsymbol{\epsilon}_0}^t \boldsymbol{R}^{-1}
\boldsymbol{z}|^2}{\boldsymbol{s}_{\boldsymbol{\epsilon}_0}^t
\boldsymbol{R}^{-1} \boldsymbol{s}_{\boldsymbol{\epsilon}_0}}.
\end{equation}

\subsection{Subpixel detectors}
\label{subsec:subpixel-detector}

Our aim is to build refined detectors that improve performance of the
above GPMF in taking into account the variability of
the object signature due to its random subpixel location. Several
solutions may be used. We first recall the most popular one.

\subsubsection{Generalized likelihood ratio test}

ML estimation of the two unknown parameters leads to the generalized
likelihood ratio test (GLRT):
\begin{equation}
\begin{array}{lcl}
\mathcal{L}_g(\boldsymbol{z})  &=& \displaystyle
\frac{\max_{(\alpha,\boldsymbol{\epsilon})} \; p(\boldsymbol{z}
|\mathrm{H_1},\alpha,\boldsymbol{\epsilon})}{p(\boldsymbol{z}
|\mathrm{H_0})} \\ \\ &=& \displaystyle \frac{p(\boldsymbol{z}
|\mathrm{H_1},\hat{\alpha}_{\mathrm{\textsc{ml}}},\hat{\boldsymbol{\epsilon}}_{\mathrm{\textsc{ml}}})}{p(\boldsymbol{z}
|\mathrm{H_0})}
\begin{array}{c}
>\\ <
\end{array}
\mathrm{threshold}.
\end{array}
\label{eq:glr}
\end{equation}
It consists in estimating the amplitude $\alpha$ \emph{and} the
possible object location $\boldsymbol{\epsilon}$ by computing:
\begin{equation}
\begin{array}{lcl}
\hat{\boldsymbol{\epsilon}}_{\mathrm{\textsc{ml}}} &=& \displaystyle
\arg \max_{\boldsymbol{\epsilon} \in \mathcal{E}} \; p(\boldsymbol{z}
|
\mathrm{H_1},\hat{\alpha}(\boldsymbol{\epsilon}),\boldsymbol{\epsilon})
\\ &=& \displaystyle \arg \max_{\boldsymbol{\epsilon} \in \mathcal{E}}
\; \left\{\frac{|\boldsymbol{s}_{\boldsymbol{\epsilon}}^t
\boldsymbol{R}^{-1}
\boldsymbol{z}|^2}{\boldsymbol{s}_{\boldsymbol{\epsilon}}^t
\boldsymbol{R}^{-1}\boldsymbol{s}_{\boldsymbol{\epsilon}}} \right\},
\end{array}
\label{eq:epsilonml}
\end{equation}
then thresholding the estimated filter
$\hat{\alpha}_{\mathrm{\textsc{ml}}}
\mathcal{T}_{\hat{\boldsymbol{\epsilon}}_{\mathrm{\textsc{ml}}}}(\boldsymbol{z})$
where  $\hat{\alpha}_{\mathrm{\textsc{ml}}} =
\hat{\alpha}(\hat{\boldsymbol{\epsilon}}_{\mathrm{\textsc{ml}}})$ is
given by equation (\ref{eq:alphaml}):
\begin{equation}
\hat{\alpha}_{\mathrm{\textsc{ml}}}
\mathcal{T}_{\hat{\boldsymbol{\epsilon}}_{\mathrm{\textsc{ml}}}}(\boldsymbol{z})
= \displaystyle
\frac{|\boldsymbol{s}_{\hat{\boldsymbol{\epsilon}}_{\mathrm{\textsc{ml}}}}^t
\boldsymbol{R}^{-1}
\boldsymbol{z}|^2}{\boldsymbol{s}_{\hat{\boldsymbol{\epsilon}}_{\mathrm{\textsc{ml}}}}^t
\boldsymbol{R}^{-1}
\boldsymbol{s}_{\hat{\boldsymbol{\epsilon}}_{\mathrm{\textsc{ml}}}}}.
\end{equation}

\subsubsection{Exact likelihood ratio test}

In a Bayesian approach, we propose to consider the two unknown
parameters $\alpha$ and $\boldsymbol{\epsilon}$ as realizations of
independent random variables with given probability density functions
$p(\alpha)$ and $p(\boldsymbol{\epsilon})$.   Then the optimal
procedure is the exact likelihood ratio test (ELRT).

To compute the density function of data under $\mathrm{H_1}$ and to get the
likelihood ratio, we have to integrate the conditional density
$p(\boldsymbol{z} | \mathrm{H_1}, \alpha, \boldsymbol{\epsilon})$ over
prior distributions of the nuisance random parameters $\alpha$
and $\boldsymbol{\epsilon}$.   The likelihood ratio can be expressed as:
\begin{equation}
\mathcal{L}(\boldsymbol{z}) = \frac{p(\boldsymbol{z}
|\mathrm{H_1})}{p(\boldsymbol{z} |\mathrm{H_0})} = \displaystyle
\frac{\int_{\mathcal{E}} \int_{\mathbbm{R}} p(\boldsymbol{z} |
\mathrm{H_1},\alpha,\boldsymbol{\epsilon}) p(\alpha)
p(\boldsymbol{\epsilon}) \,d\alpha \,d\boldsymbol{\epsilon}}
{p(\boldsymbol{z} | \mathrm{H_0})}.
\label{eq:lr}
\end{equation}
Given prior distributions $p(\alpha)$ and $p(\boldsymbol{\epsilon})$,
$\mathcal{L}(\boldsymbol{z})$ is the optimal Neyman-Pearson test
whenever $\alpha$ and $\boldsymbol{\epsilon}$ really satisfy the
models $p(\alpha)$ and $p(\boldsymbol{\epsilon})$.   By default we
choose a ``\,non-informative\,'' prior for $\alpha$ and we adopt a
uniform distribution inside the pixel for $\boldsymbol{\epsilon}$,
which seems to be quite a reasonable assumption for the subpixel
target position.   So we get:
\begin{equation}
\mathcal{L}(\boldsymbol{z}) \propto \int_{\mathcal{E}}
\frac{1}{\sqrt{\boldsymbol{s}_{\boldsymbol{\epsilon}}^t
\boldsymbol{R}^{-1}\boldsymbol{s}_{\boldsymbol{\epsilon}}}}  \exp
\left\{  \frac{|\boldsymbol{s}_{\boldsymbol{\epsilon}}^t
\boldsymbol{R}^{-1} \boldsymbol{z}|^2} {2 \,
\boldsymbol{s}_{\boldsymbol{\epsilon}}^t \boldsymbol{R}^{-1}
\boldsymbol{s}_{\boldsymbol{\epsilon}}} \right\}
\,d\boldsymbol{\epsilon}.
\label{eq:lrprop}
\end{equation}

Unfortunately, because of intricate nonlinear dependence of
$\boldsymbol{s}_{\boldsymbol{\epsilon}}$ on $\boldsymbol{\epsilon}$,
explicit integration over $\boldsymbol{\epsilon}$ appears to be not
tractable and probability distribution of
$\mathcal{L}(\boldsymbol{z})$ is not as simple as the one of
$\mathcal{T}_{\boldsymbol{\epsilon}_0}(\boldsymbol{z})$.  
A quadrature approximation is required to compute
$\mathcal{L}(\boldsymbol{z})$ whereas derivation of its density
requires Monte-Carlo simulations.

\subsubsection{Approximate likelihood ratio test}

In equation (\ref{eq:lrprop}), the double integral over
$\boldsymbol{\epsilon}$ can be approximated up to any desired accuracy
using some quadrature rule and evaluating the integrand
$f(\boldsymbol{\epsilon} | \boldsymbol{z})$ at discrete samples
$\boldsymbol{\epsilon}_k \in \mathcal{E}=[-0.5,0.5[^2$. But, for sake
of computational efficiency, we propose to use a coarsest
approximation of the likelihood ratio (ALRT) based on a bidimensional trapezoidal
rule which only involves the $9$ half-pixel positions.

\subsubsection{Subspace model}

One alternative to this probabilistic viewpoint can be built on a
geometric approach that restricts the signal vector
$\boldsymbol{s}=\alpha \boldsymbol{s}_{\boldsymbol{\epsilon}}$ to vary
in some $P$-dimensional subspace, with $P$ lower than the vector size
\cite{Manolakis02}.   The observed data under $\mathrm{H_1}$ are
rewritten as:
\begin{equation}
\boldsymbol{z} \simeq \boldsymbol{S}
\boldsymbol{a}+\boldsymbol{n}=\sum_{p=1}^P a_p \boldsymbol{s}_p +
\boldsymbol{n},
\end{equation}
where the structural matrix $\boldsymbol{S}$ is formed by $P$
independent vectors $\boldsymbol{s}_p$. Coefficients $a_p$ of the
linear combination are the new parameters that describe the signal
variability. Thanks to linearity, ML estimation of vector
$\boldsymbol{a}$ has an explicit solution (which is identical to the
least squares estimator):
\begin{equation}
\hat{\boldsymbol{a}}_{\mathrm{\textsc{ml}}}=(\boldsymbol{S}^t
\boldsymbol{R}^{-1} \boldsymbol{S})^{-1} \boldsymbol{S}^t
\boldsymbol{R}^{-1} \boldsymbol{z}
\end{equation}
and GLRT amounts to threshold the following statistic:
\begin{equation}
\mathcal{D}(\boldsymbol{z})=\boldsymbol{z}^t \boldsymbol{R}^{-1}
\boldsymbol{S} (\boldsymbol{S}^t \boldsymbol{R}^{-1}
\boldsymbol{S})^{-1} \boldsymbol{S}^t \boldsymbol{R}^{-1}
\boldsymbol{z}.
\end{equation}
Matrix $\boldsymbol{S}$ only depends on $\boldsymbol{\epsilon}$,
$\alpha$ being a scale parameter.   In practice, it is identified by
discretizing $\mathcal{E}$, making a singular value decomposition and
retaining the singular vectors $\boldsymbol{s}_p$ corresponding to the
$P$ greatest singular values.   We choose $P=1$ which gives better
results than higher orders. Therefore under hypothesis $\mathrm{H_1}$,
$\boldsymbol{z} \simeq a_1 \boldsymbol{s}_1 + \boldsymbol{n}$ and
$\mathcal{D}(\boldsymbol{z})$ is identical to GPMF with
$\boldsymbol{s}_1$ replacing
$\boldsymbol{s}_{\boldsymbol{\epsilon}_0}$.

\section{Application to optical imagery}
\label{sec:application}

\subsection{Optical system}

\begin{figure}[ht]
\begin{center}
\begin{tabular}[]{cc}
\includegraphics[height=2.5cm]{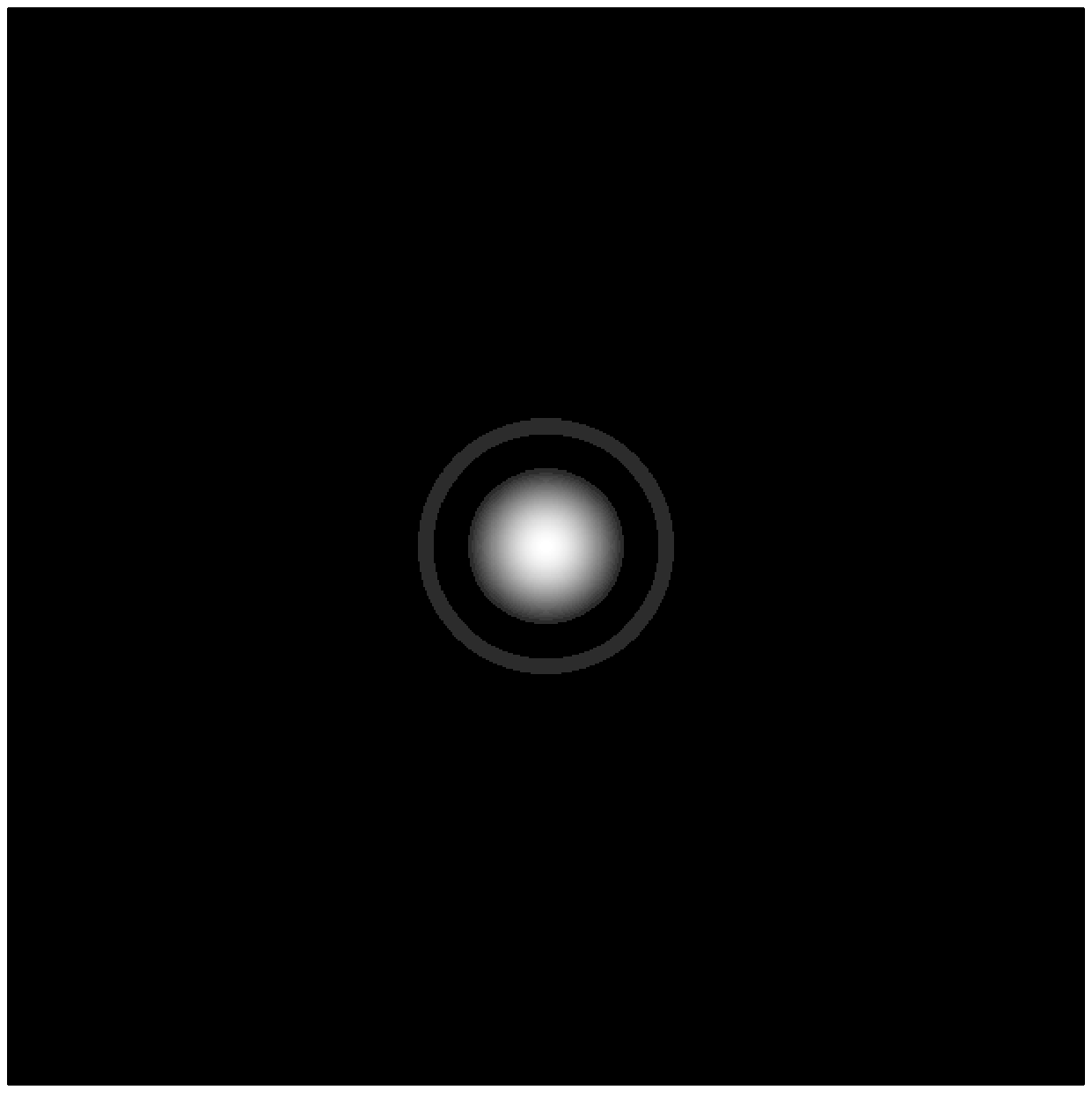}
&\includegraphics[height=2.5cm]{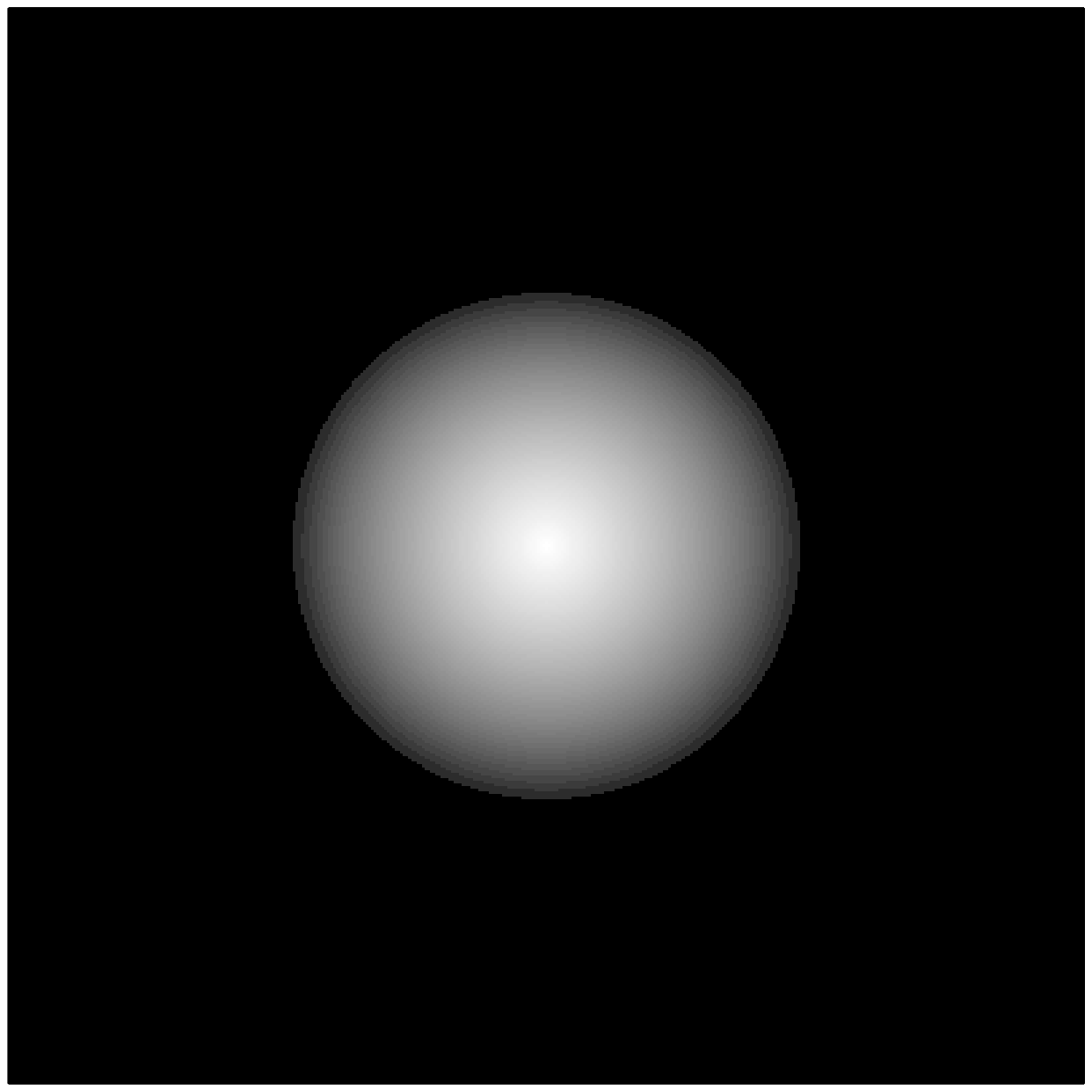}\\
\includegraphics[height=2.5cm]{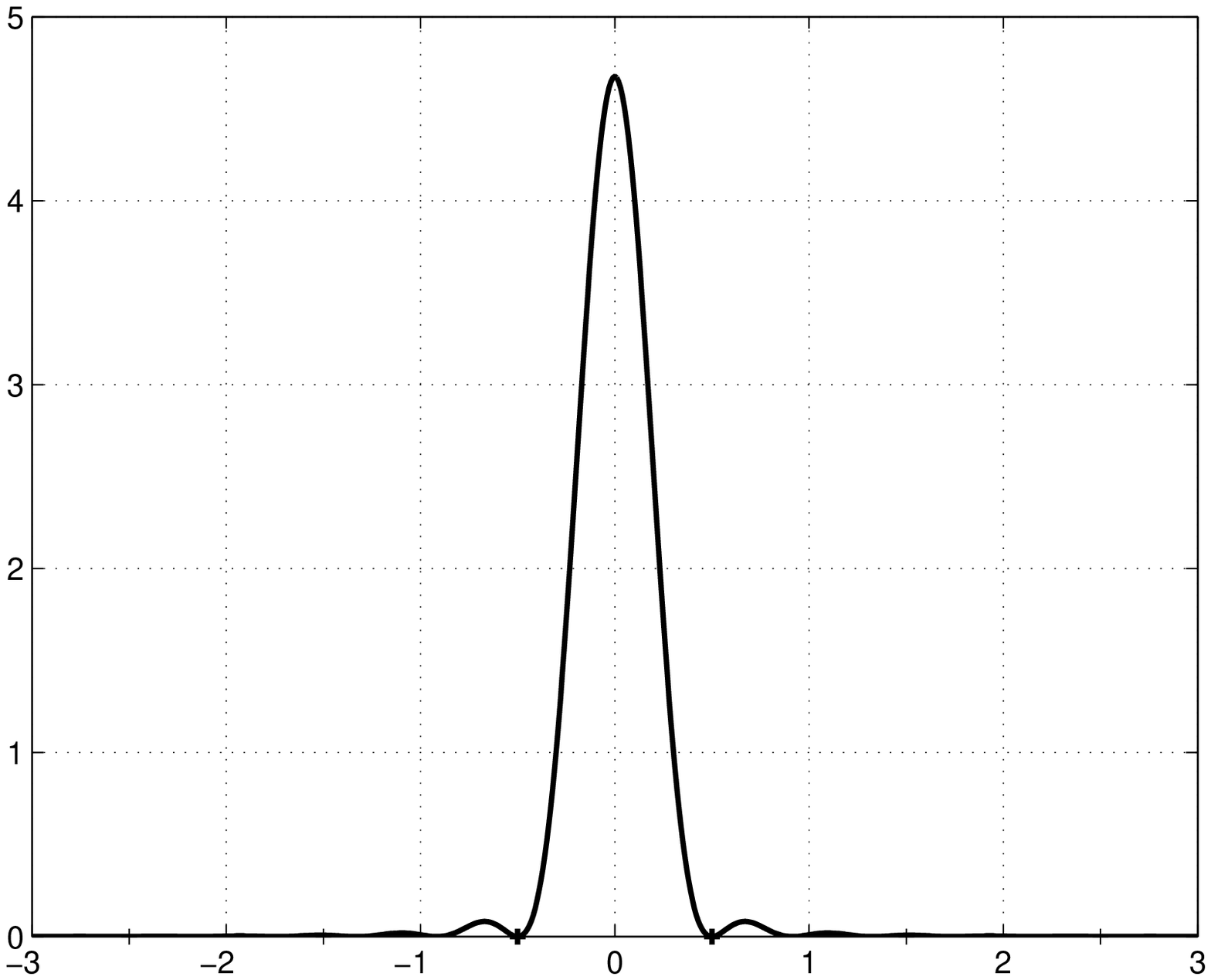}
&\includegraphics[height=2.5cm]{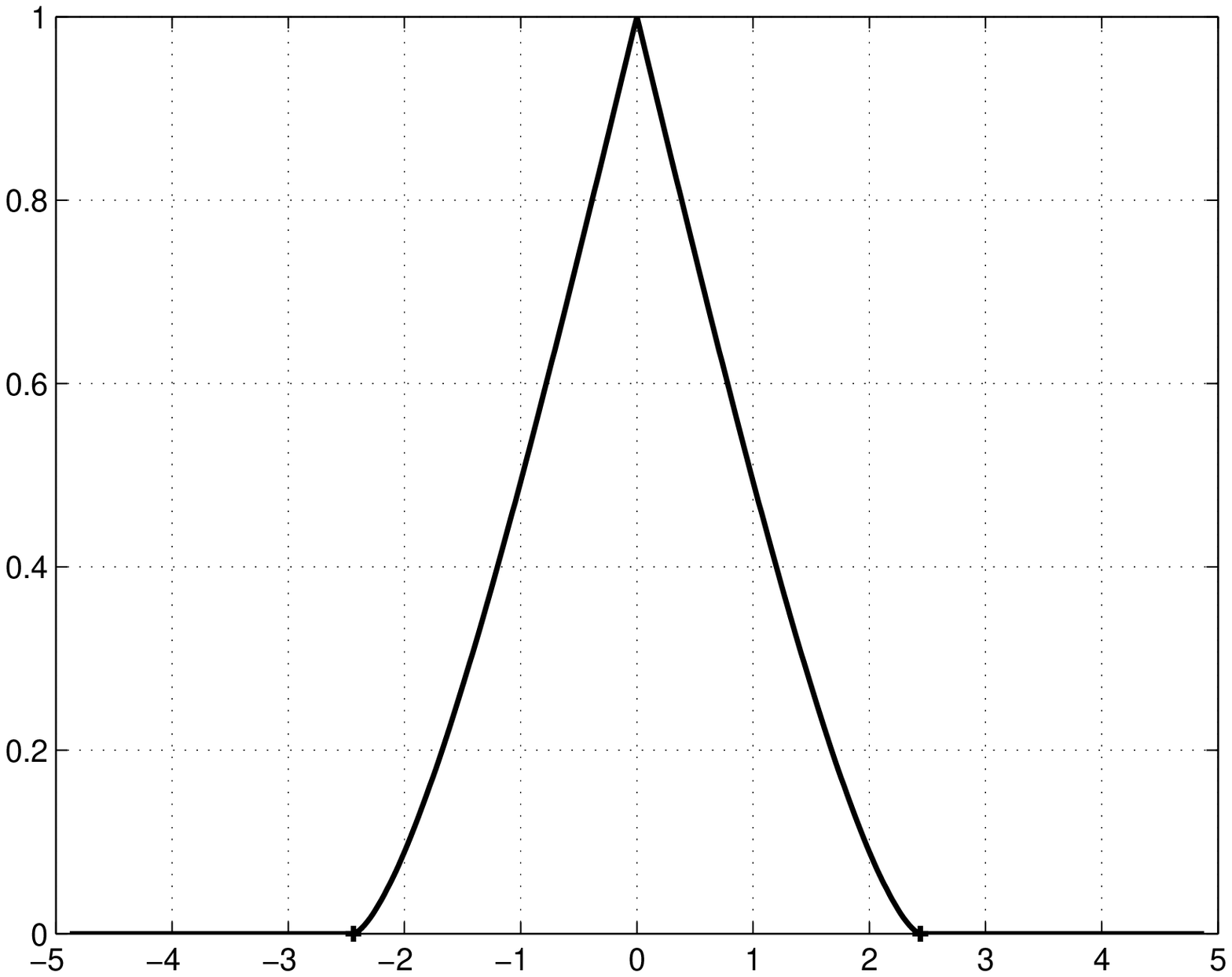} \\ $\rho$ & $r$
\end{tabular}
\caption[]{\emph{Left}: radial point spread function $h_o(u,v)$ on the top and
slice along a diameter on the bottom. \emph{Right}: corresponding
optical transfer function $\tilde{h}_o(\nu_u,\nu_v)$ and slice along a
diameter ($r_c=2.44$).}
\label{fig:psf}
\end{center}
\end{figure}

In our application, we can model the imaging system by a
diffraction-limited, unaberrated optics with circular aperture and
incoherent illumination \cite{Goodman72,Hardie98}.  The object signal
pattern $\boldsymbol{s}_{\boldsymbol{\epsilon}}$  is then given by the
integration of $h_o$ on each pixel (see equation~\ref{eq:signal}),
where $h_o$ is the radial point spread function (PSF) defined by the
Airy disk:
\begin{equation}
h_o(u,v)=\frac{1}{\pi} \left[\frac{J_1(\pi \rho \, r_c)}{\rho}
\right]^2, \quad \rho=\sqrt{u^2+v^2}.
\label{eq:ho}
\end{equation}
$J_1$ is the Bessel function of the first kind and
$r_c={\nu_c}/{\nu_s}$ designates the normalized cut-off frequency
($\nu_s$ is the sampling frequency and $\nu_c={D}/{\lambda}$ is the
radial cut-off frequency defined by the ratio of the lens aperture
diameter $D$ over the wavelength $\lambda$).    
Figure~\ref{fig:psf} depicts the two-dimensional PSF and a
slice along one diameter, as well as their Fourier transform.   Common
sensor design uses $r_c=2.44$ so that the pixel size is equal to the
width of the main lobe of the PSF.   However, this implies a
downsampling factor of ${\nu_n}/{\nu_s}=2 \, r_c=4.88$ (where $\nu_n=2
\, \nu_c$ is the Nyquist frequency).   In the following section, we
present some numerical results of detection performance considering
this classical sensor design.   Examples of image spots
$\boldsymbol{s}_{\boldsymbol{\epsilon}}$ have been represented on
Figure~\ref{fig:taches} for different values of
$\boldsymbol{\epsilon}$.

\begin{remark}
We have the following property:
$$\sum_{(i,j) \in \mathbbm{Z}^2}
\boldsymbol{s}_{\boldsymbol{\epsilon}}[i,j] = \int_{\mathbbm{R}^2}
h_o(u,v) \,du \,dv = 1.$$
\end{remark}

\subsection{Numerical results}

\begin{figure}[ht]
\begin{center}
\begin{tabular}[]{c}
\includegraphics[height=4.5cm]{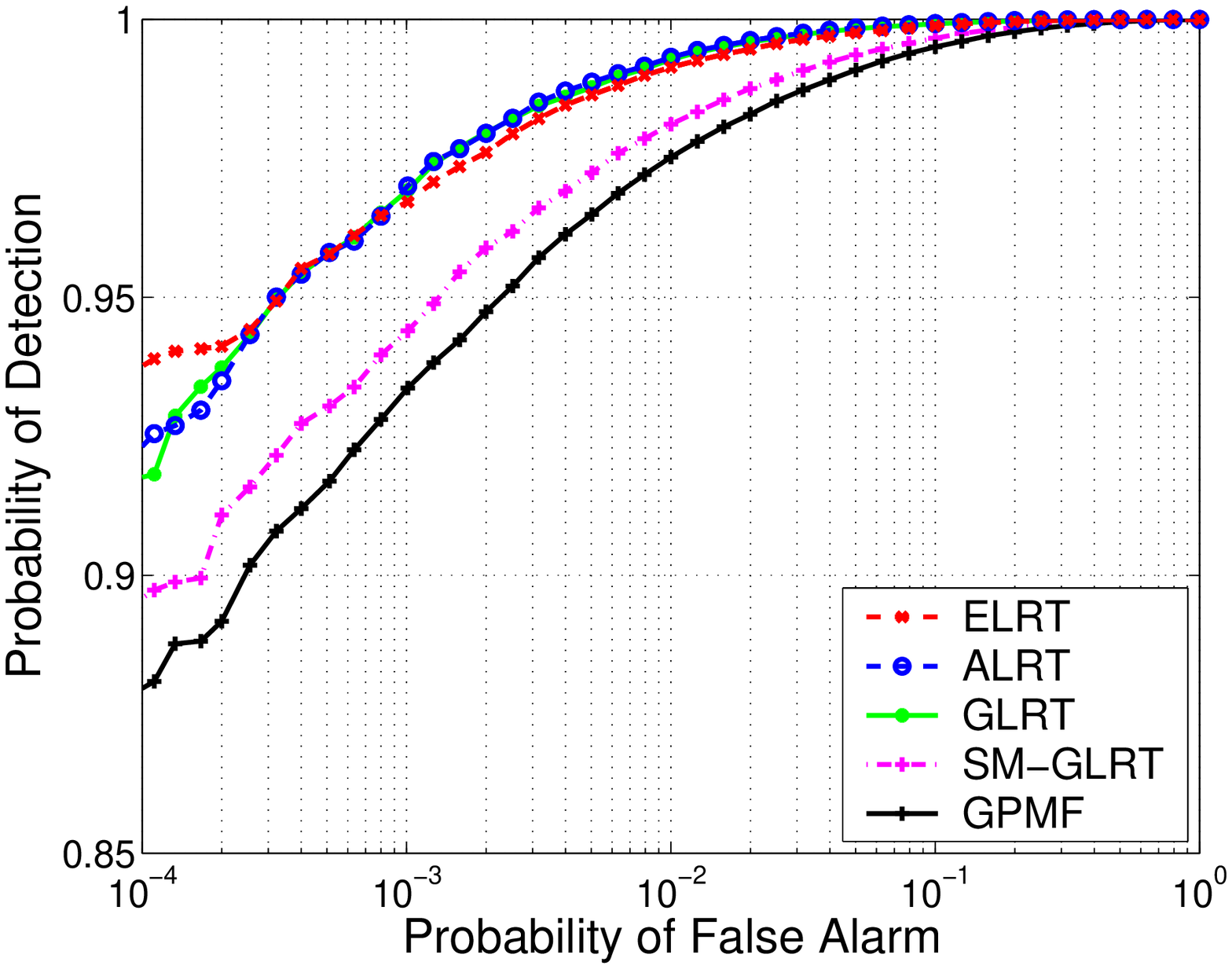} \\ SNR $=16.2$dB ($\alpha /
\sigma = 9$) \\ \\ \includegraphics[height=4.5cm]{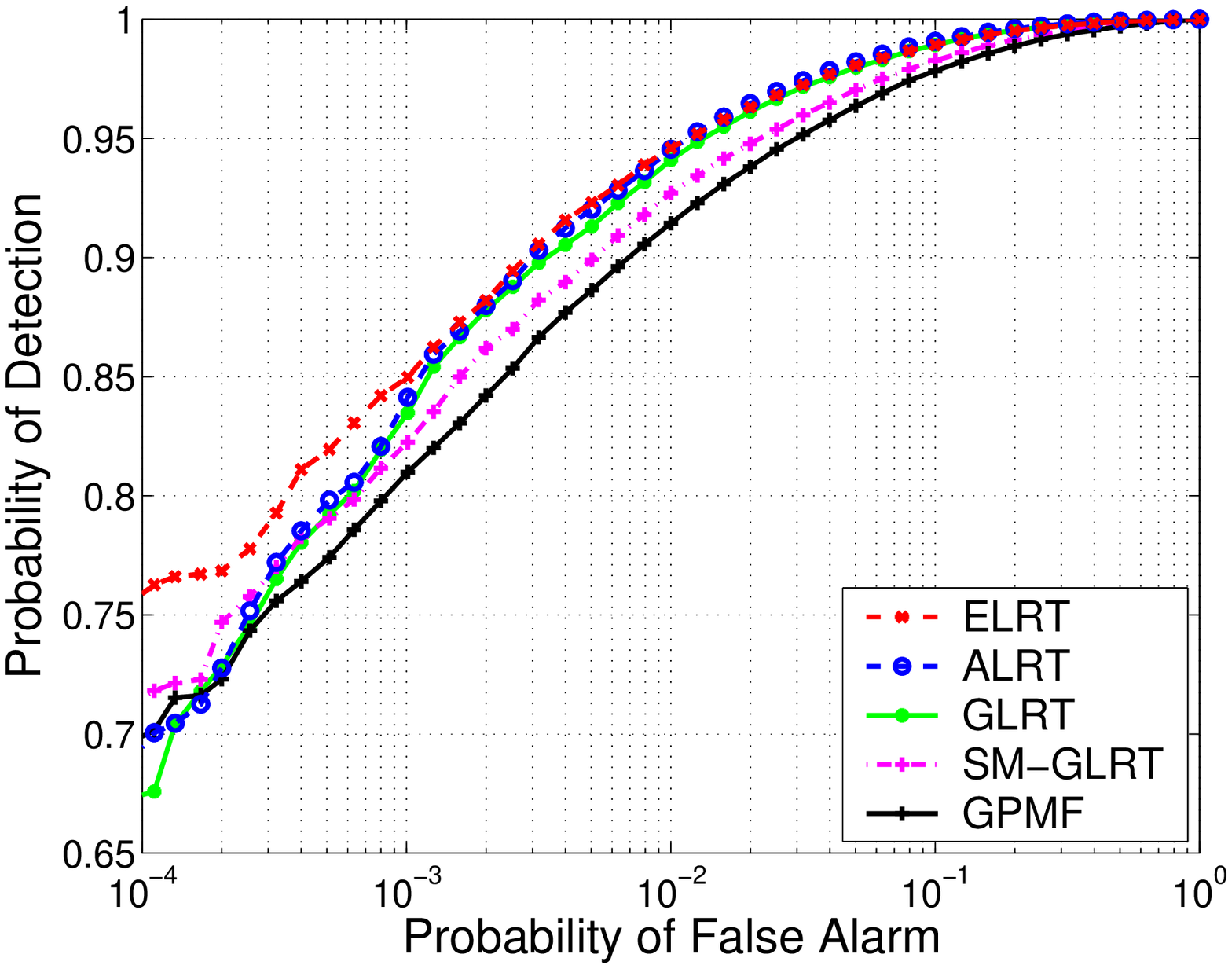} \\ SNR
$=14.1$dB ($\alpha / \sigma = 7$)
\end{tabular}
\caption[]{Empirical ROC curves in the Gaussian white noise case with common sensor design ($r_c=2.44$).}
\label{fig:bb7090}
\end{center}
\end{figure}

Performances of the five classes of detectors have been compared in
terms of ROC curves: GPMF, GLRT on $\alpha$ and
$\boldsymbol{\epsilon}$, ELRT, ALRT and finally GLRT with the subspace
model denoted SM-GLRT.   Probabilities of detection and false alarm
are deduced from the empirical distributions of these statistics under
each hypothesis by generating samples of Gaussian noise
$\boldsymbol{n}$ and uniformly distributed $\boldsymbol{\epsilon}$ in
$\mathcal{E}=[-0.5,0.5[^2$.   The amplitude is assumed to be unknown
but set to a constant value $\alpha$ in the simulations since we
have no information about a reliable prior distribution $p(\alpha)$.

We first consider the case of a Gaussian white noise $\boldsymbol{n}
\sim \mathcal{N}(0,\sigma^2)$.   
The signal-to-noise ratio is then defined by:
\begin{eqnarray}
&& \mathrm{SNR} = 10 \, \log_{10} \left(\frac{\alpha^2 E}{\sigma^2}
\right) \\ &\mathrm{with}& E=\int_{\mathcal{E}} \sum_{(i,j) \in
\mathbbm{Z}^2} \left(\boldsymbol{s}_{\boldsymbol{\epsilon}}[i,j]
\right)^2 \,d\boldsymbol{\epsilon} \nonumber
\label{eq:snr}
\end{eqnarray}
For common sensor design ($r_c=2.44$), the average energy of the
image spot is $E \simeq 0.52$.   The ROC curves are depicted on
Figure~\ref{fig:bb7090} for two different SNR.  It shows that the
GLRT, the ELRT (actually, a refined approximation of it) and the
coarse approximation ALRT give significantly better performance than
the SM-GLRT and GPMF.    We also see that the performance gain is
greater for high SNR whereas it tends to be rather small for low SNR
and low probability of false alarm.   Conversely, if the latter
detectors are computationally cheap, including the ALRT, this is not
so for the GLRT and the ELRT, which are much more intensive.

\begin{figure}[ht]
\begin{center}
\includegraphics[height=4.5cm]{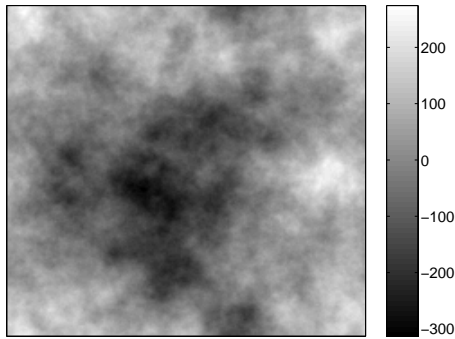}
\caption[]{Simulation of a cloud fractal image of $200 \times 200$ pixels (Hurst parameter $H=0.7$).}
\label{fig:imagefbm}
\end{center}
\end{figure}

\begin{figure}[ht]
\begin{center}
\includegraphics[height=4.5cm]{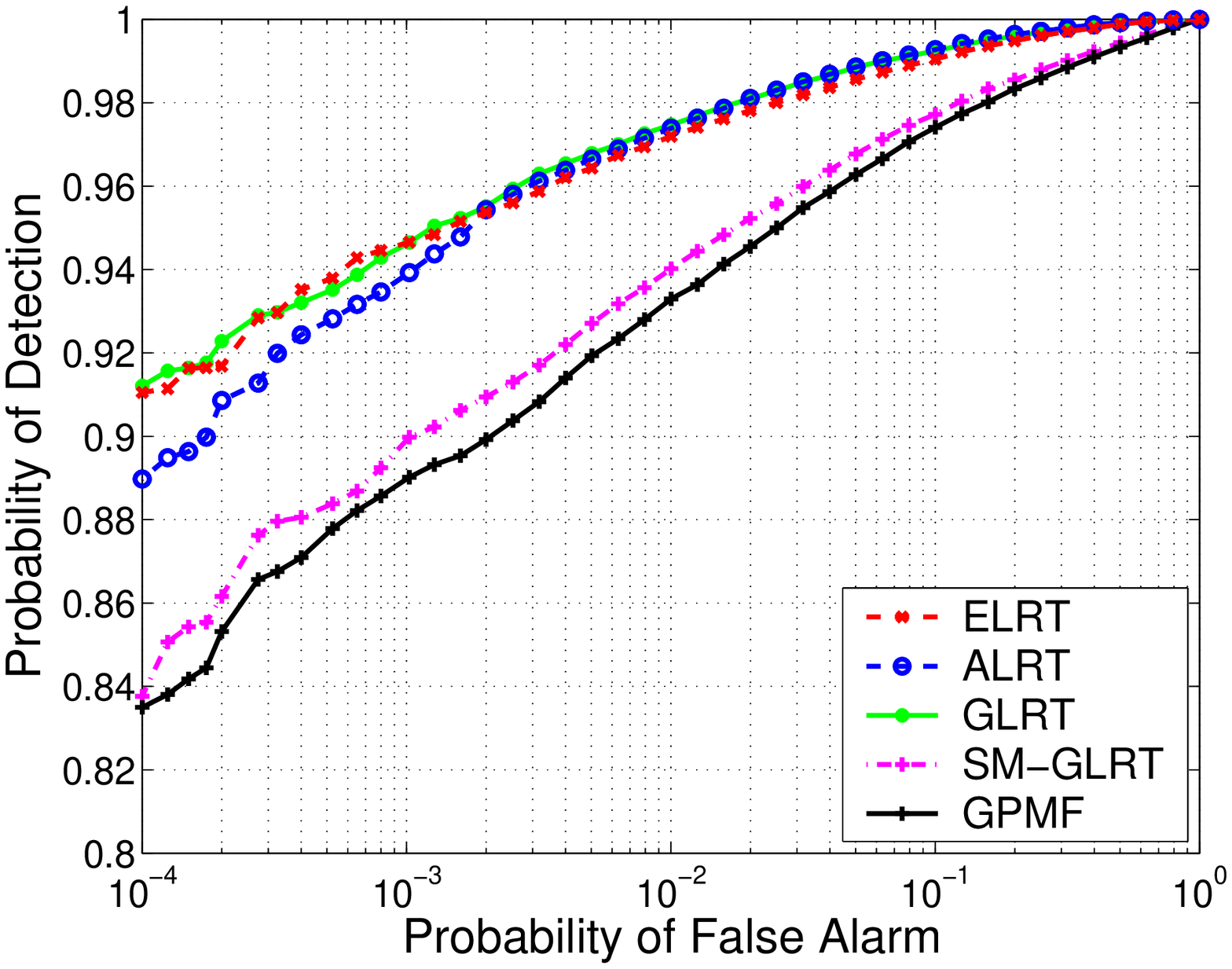}
\caption[]{Empirical ROC curves obtained on the fractal image of Figure~\ref{fig:imagefbm} for a true (but assumed unknown) target amplitude $\alpha=60$ gray levels.}
\label{fig:fbm60}
\end{center}
\end{figure}

As complementary results, we have tested the five detectors on a
fractal background image generated by a variant of the ppmforge
software
\footnote{http://h30097.www3.hp.com/demos/ossc/man-html/man1/ppmforge.1.html}.
The synthesis algorithm depends on the auto-similarity parameter $H$
called Hurst parameter and which is set to $0.7$ in this experiment.
The resulting image depicted on Figure~\ref{fig:imagefbm} is a
realistic simulation of a cloud scene.   The covariance matrix
$\boldsymbol{R}$ of this stationary background is estimated by the
empirical correlations on the whole image. We then compute the
performance of the different detectors for a given target amplitude as
illustrated on Figure~\ref{fig:fbm60}.   The ROC curves look quite
different from the white noise case but we notice again that the GLRT,
ELRT and ALRT have similar performance and provide a significant
detection gain in comparison with the GPMF or the SM-GLRT.


\subsection{Influence of the optics}
\label{subsec:influenceofoptics}

\begin{figure}[ht]
\begin{center}
\includegraphics[height=4.5cm]{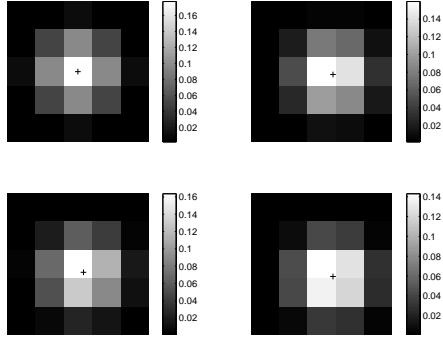}
\caption[]{Examples of image spots corresponding to a correctly sampled optics ($r_c=0.5$) to be compared to those of Figure~\ref{fig:taches}.}
\label{fig:taches05}
\end{center}
\end{figure}

\begin{figure*}[ht]
\begin{center}
\setlength{\tabcolsep}{1cm}
\begin{tabular}[]{cc}
\includegraphics[height=4.5cm]{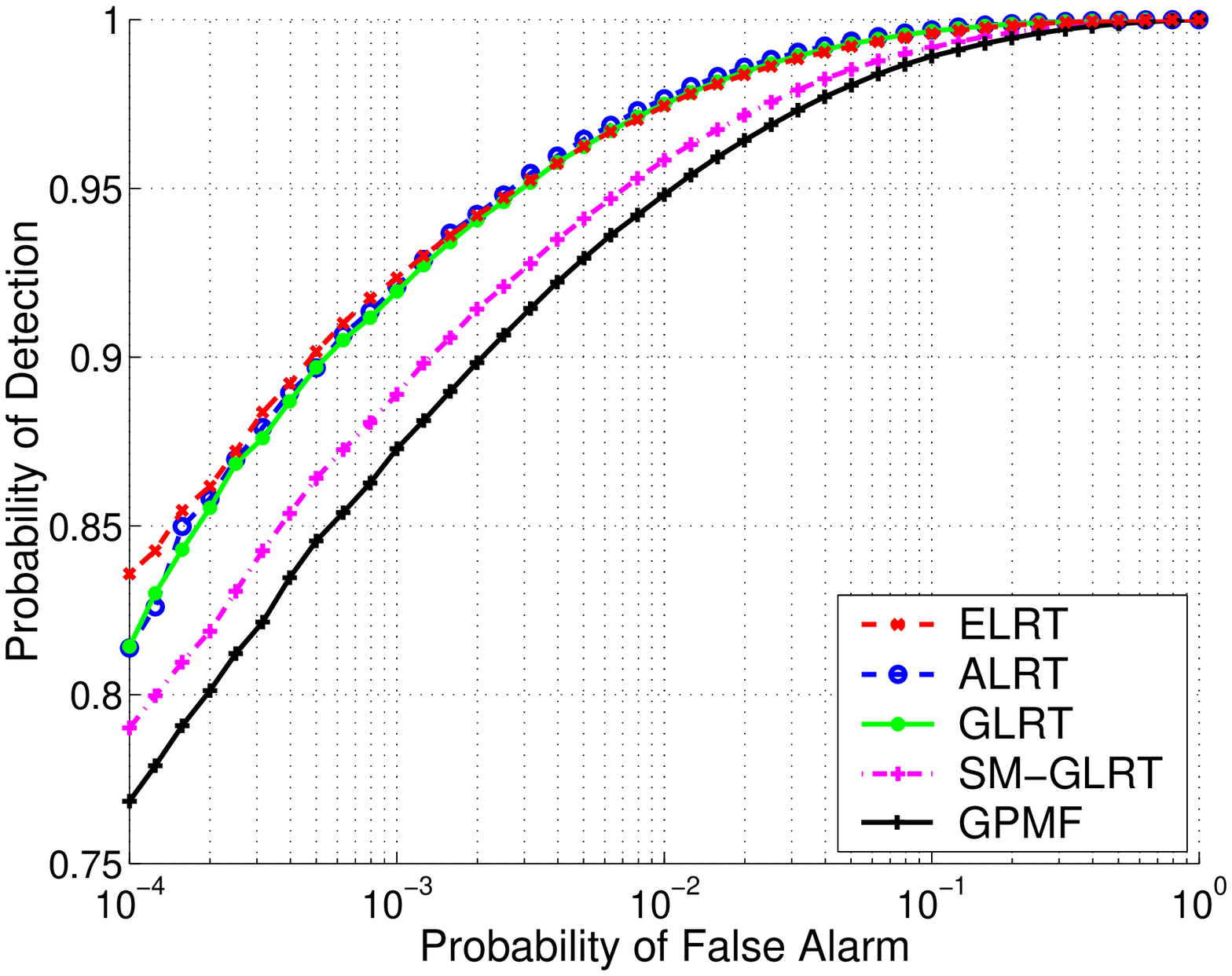} &
\includegraphics[height=4.5cm]{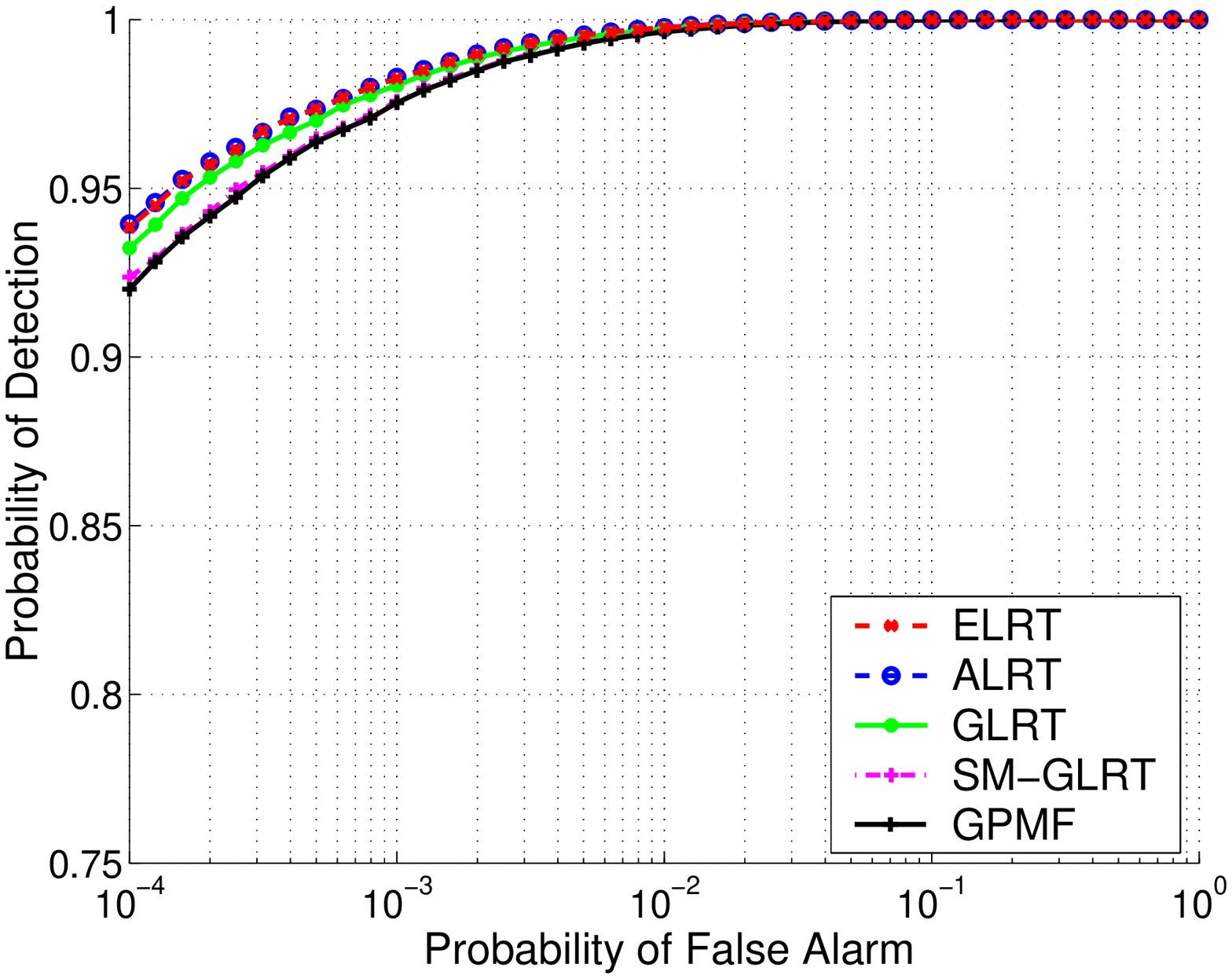} \\ $r_c=2.44$ &
$r_c=0.5$
\end{tabular}
\caption[]{Empirical ROC curves in the Gaussian white noise case for a same SNR $=15$dB,
with common sensor design on the left ($r_c=2.44$) compared to a
correctly sampled optics on the right ($r_c=0.5$).}
\label{fig:cor}
\end{center}
\end{figure*}

Besides the perfecting and evaluation of subpixel detectors, one
additional motivation of this work is to analyze the influence of
aliasing on detection performance.   This is the reason why we have
also tested the detectors on a correctly sampled optics in order to
compare their performances with those obtained using a common sensor
design.   In the correctly sampled design, the focal plane is sampled at Nyquist
frequency (implying a denser sensor array or a smaller lens diameter)
so that aliasing is suppressed. Parameter $r_c$ of the PSF is
equal to $0.5$ and the signal energy is now spread over several pixels ($E
\simeq 0.08$).   By comparison, Figure~\ref{fig:taches05} presents
the examples of image spots corresponding to such a design.
Detection performances are depicted on Figure~\ref{fig:cor} on the
right for a SNR of $15$dB.   We see that improved detection has just a moderate impact in
this situation.   The five detectors have a quite similar behavior
but at the same SNR they perform much better than in the aliased
case. The gain in $P_{fa}$ amounts at least to a
factor $10$ for all the detectors. Such a result speaks in favour of
using a denser focal plane for point target detection.

\section{Performance of subpixel position estimators}
\label{sec:position-estimation}

Up to now we have focused on the detection strategy.   In a second
step, once a potential target is detected on a given pixel, we are
also interested in accurate estimation of its subpixel position.
Such a problem has already been addressed, in particular for 
star position estimation in astronomical applications \cite{Winick86}.
Several types of estimators are possible. We consider here the maximum
likelihood (ML) estimator and following the Bayesian approach introduced previously 
the posterior mean (PM).   It is important to note that the signal amplitude $\alpha$ is
also unknown and therefore we have to estimate it or integrate over it.
Indeed it is not valid to suppose that the amplitude is known in the context of IRST.

The ML estimator of $\boldsymbol{\epsilon}$ is given in equation (\ref{eq:epsilonml}) by
replacing $\alpha$ with its estimate $\hat{\boldsymbol{\epsilon}}$.   Actually,
$\hat{\boldsymbol{\epsilon}}_{\mathrm{\textsc{ml}}}$ and
$\hat{\alpha}_{\mathrm{\textsc{ml}}} =
\hat{\alpha}(\hat{\boldsymbol{\epsilon}}_{\mathrm{\textsc{ml}}})$ are
identical to joint maximum \textit{a posteriori} (MAP) estimators with
non-informative priors on the two parameters.

The PM estimator is defined as:
\begin{equation}
\hat{\boldsymbol{\epsilon}}_{\mathrm{\textsc{pm}}} =
\int_{\mathcal{E}} \boldsymbol{\epsilon} \, p(\boldsymbol{\epsilon} |
\mathrm{H_1},\boldsymbol{z}) \, d\boldsymbol{\epsilon}
\label{eq:epsilonpm}
\end{equation}
where the posterior law is deduced from Bayes'rule:
\begin{eqnarray}
p(\boldsymbol{\epsilon} | \mathrm{H_1},\boldsymbol{z})
&=&\displaystyle \frac{ p(\boldsymbol{z} |
\mathrm{H_1},\boldsymbol{\epsilon}) \, p(\boldsymbol{\epsilon}) }{
p(\boldsymbol{z} | \mathrm{H_1}) } \\ &=& \displaystyle \frac{
p(\boldsymbol{\epsilon}) }{ p(\boldsymbol{z} | \mathrm{H_1}) } \,
\int_{\mathbbm{R}} p(\boldsymbol{z} |
\mathrm{H_1},\alpha,\boldsymbol{\epsilon}) \, p(\alpha) \,
d\alpha. \nonumber
\label{eq:loiapostepsilon}
\end{eqnarray}
So, we have to integrate over $\alpha$ and then over
$\boldsymbol{\epsilon}$. As previously we consider a diffuse \textit{a
priori} on $\mathbbm{R}$ for $\alpha$ and a uniform law on
$\mathcal{E}$ for $\boldsymbol{\epsilon}$. We get the following expression in the same way
as for the likelihood ratio in equation (\ref{eq:lrprop}):
\begin{equation}
p(\boldsymbol{\epsilon} | \mathrm{H_1},\boldsymbol{z}) \propto \frac{
1 }{ \sqrt{\boldsymbol{s}_{\boldsymbol{\epsilon}}^t
\boldsymbol{R}^{-1} \boldsymbol{s}_{\boldsymbol{\epsilon}} } } \exp
\left\{ \frac{ |\boldsymbol{s}_{\boldsymbol{\epsilon}}^t
\boldsymbol{R}^{-1} \boldsymbol{z}|^2 }{ 2 \,
\boldsymbol{s}_{\boldsymbol{\epsilon}}^t \boldsymbol{R}^{-1}
\boldsymbol{s}_{\boldsymbol{\epsilon}} } \right\}.
\label{eq:loiapost2}
\end{equation}

\begin{figure*}[ht]
\begin{center}
\setlength{\tabcolsep}{1cm}
\begin{tabular}[]{cc}
\includegraphics[height=4.5cm]{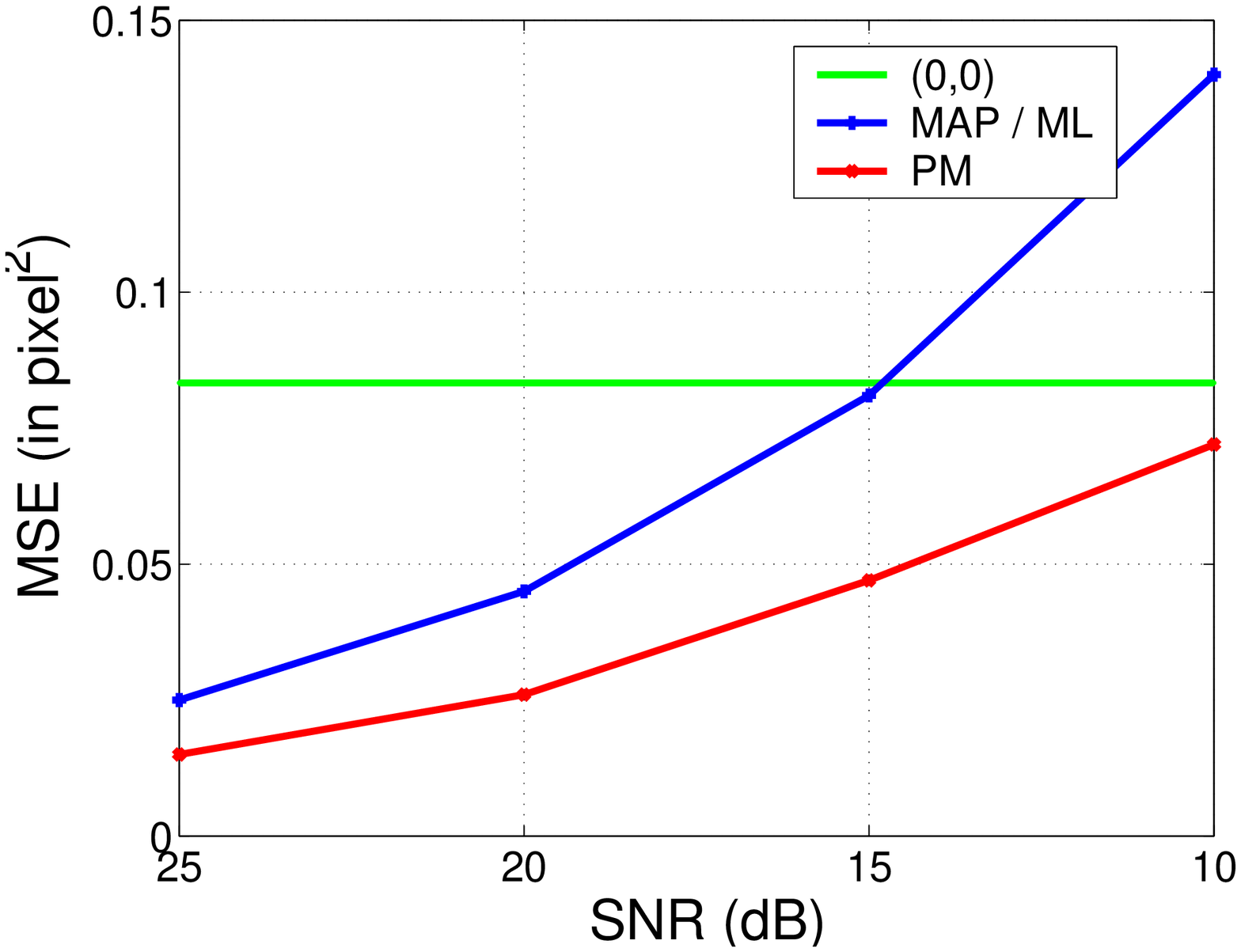} &
\includegraphics[height=4.5cm]{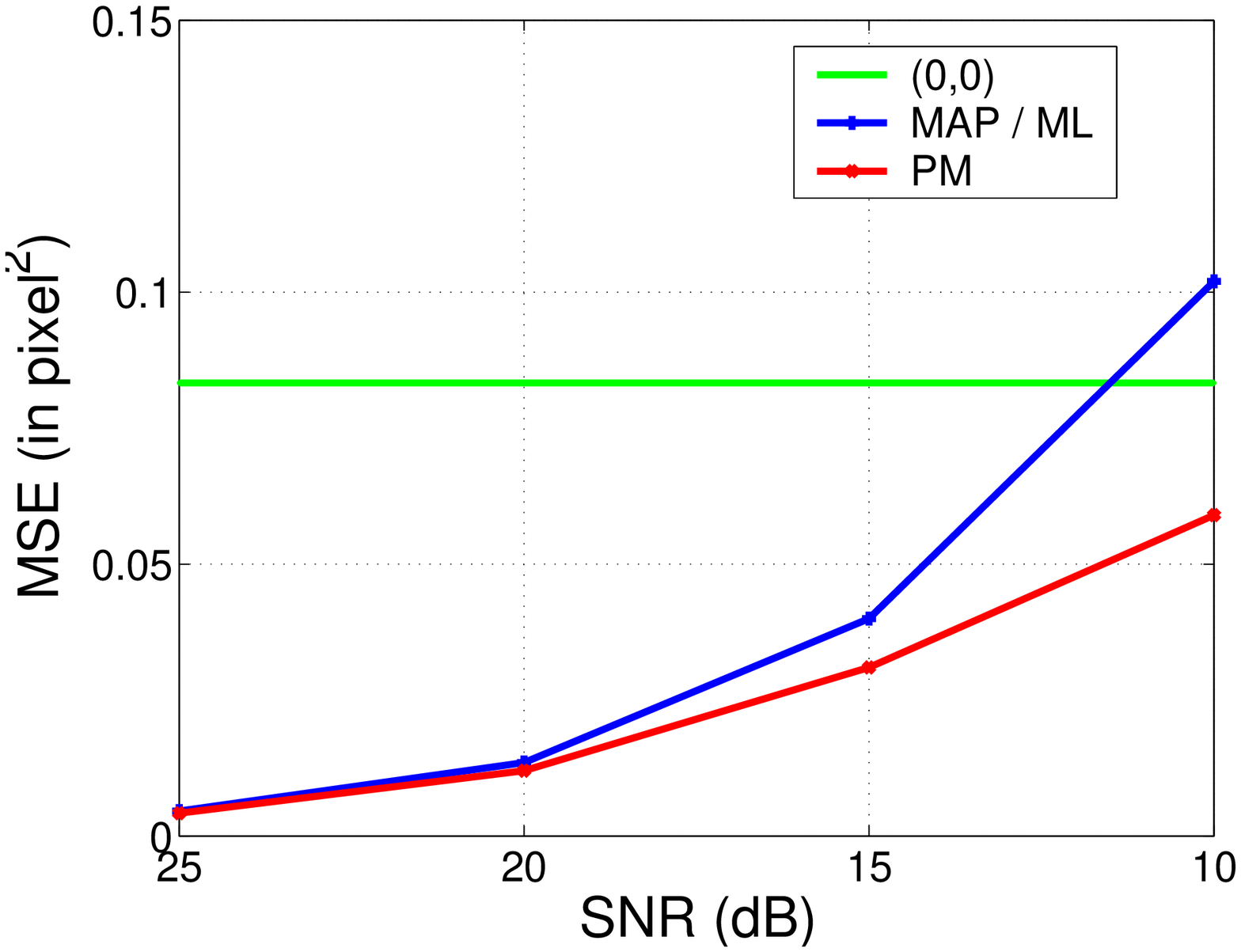} \\ $r_c=2.44$ &
$r_c=0.5$
\end{tabular}
\caption[]{Average mean square errors (MSE) of position estimators in the Gaussian white noise case with common sensor design on the left ($r_c=2.44$) compared to a correctly sampled optics on the right ($r_c=0.5$).}
\label{fig:eqm}
\end{center}
\end{figure*}

We have studied the performance of these two estimators in terms of
average mean square error (MSE). In practice, the optimization or the
integration over $\boldsymbol{\epsilon}$ are approximated numerically
by considering a finite discrete grid of $20 \times 20$ values
$\boldsymbol{\epsilon}_k \in \mathcal{E}$.   Given a true position
$\boldsymbol{\epsilon}^{\star}$, bias and variance of an
estimator $\hat{\boldsymbol{\epsilon}}$ are estimated thanks to
Monte-Carlo simulations.   We consider the case of a Gaussian white
noise and we vary the signal-to-noise ratio.   Figure~\ref{fig:eqm} on
the left compares ML and PM estimators to the pixel estimator which
assumes by default that the target location is at the center of the
pixel ($\hat{\boldsymbol{\epsilon}}=(0,0)$) and whose MSE is equal to
$1/12$.   At favorable SNR, the two subpixel estimators are far better
than the default estimator but the gain decreases when the noise
becomes important.   For a SNR of $15$dB, the ML yields an error
similar to the default estimator while the PM notably has a twice smaller
error.   By comparison, Figure~\ref{fig:eqm} on the right shows the
estimation performances obtained in the unaliased case ($r_c=0.5$) for equivalent
signal-to-noise ratios. ML and PM logically perform better since the
signal is correctly sampled.

\section{Conclusion and future work}
\label{sec:conclusion}

We have presented the detection problem of subpixel objects embedded
in additive Gaussian noise.   Subpixel location and signal amplitude are assumed to be
unknown. Unknown subpixel location has a great influence 
on detection performance in the aliased case  while
conventional matched filter neglects it.   Thus, we derived four types
of improved detectors from the likelihood ratio: the GLRT, the ELRT,
the ALRT and the SM-GLRT.   We have illustrated their performance in
comparison with the more classical GPMF.   Numerical results for both
white and correlated noise cases show that the ELRT, the ALRT and the
GLRT are competitive whereas the SM-GLRT does not reach the same
quality  but slightly improves the performance of the GPMF too.
The ALRT seems to be a good trade-off since it is not as
computionnally demanding as the ELRT and the GLRT. Moreover the
performance gain  proves to be only moderate in the case of unaliased
optics.   This conclusion has important consequence in sensor design:
it suggests that  the popular design of a pixel covering exactly the
main lobe of the Airy disk is not optimum for point object detection.
Future work consists in studying the robustness of
these detectors to real data and the way we can take into
account non Gaussian distributions of background noise.   As far as the position
estimation problem is concerned, we have demonstrated prospective
gains that must also be confirmed on more realistic data.

\bibliographystyle{ieeetr}


\end{document}